\documentclass{webofc}
\usepackage[varg]{txfonts}   
\usepackage{listings}
\usepackage{xcolor}

\definecolor{codegreen}{rgb}{0,0.6,0}
\definecolor{codegray}{rgb}{0.5,0.5,0.5}
\definecolor{codepurple}{rgb}{0.58,0,0.82}
\definecolor{backcolour}{rgb}{0.95, 0.95, 0.96}

\lstdefinestyle{mystyle}{
    backgroundcolor=\color{backcolour},   
    commentstyle=\color{codegreen},
    keywordstyle=\color{magenta},
    numberstyle=\tiny\color{codegray},
    stringstyle=\color{codepurple},
    basicstyle=\ttfamily\footnotesize,
    breakatwhitespace=false,         
    breaklines=true,                 
    captionpos=b,                    
    keepspaces=true,                 
    numbers=left,                    
    numbersep=5pt,                  
    showspaces=false,                
    showstringspaces=false,
    showtabs=false,                  
    tabsize=2
}

\lstdefinestyle{interfaces}{
  float=bp,
  floatplacement=bp,
  abovecaptionskip=-5pt
}
\begin{document}
%
\title{Towards Real-World Applications of ServiceX, an Analysis Data Transformation System}
%
%

\author{\firstname{K.} \lastname{Choi}\inst{1}\fnsep\thanks{\email{kyungeonchoi@utexas.edu}}         \and
        \firstname{A.} \lastname{Eckart}\inst{2}
        \and
        \firstname{B.} \lastname{Galewsky}\inst{3}
        \and
        \firstname{R.} \lastname{Gardner}\inst{2}
        \and
        \firstname{M.} \lastname{Neubauer}\inst{3}
        \and
        \firstname{P.} \lastname{Onyisi}\inst{1}
        \and
        \firstname{M.} \lastname{Proffitt}\inst{4}
        \and
        \firstname{I.} \lastname{Vukotic}\inst{2}
        \and
        \firstname{G.} \lastname{Watts}\inst{4}
}

\institute{University of Texas at Austin \and The University of Chicago \and University of Illinois at Urbana-Champaign \and University of Washington}

\abstract{%
    One of the biggest challenges in the High-Luminosity LHC (HL-LHC) era will be the significantly increased data size to be recorded and analyzed from the collisions at the ATLAS and CMS experiments. ServiceX is a software R$\&$D project in the area of Data Organization, Management and Access of the IRIS- HEP to investigate new computational models for the HL-LHC era. 
    ServiceX is an experiment-agnostic service to enable on-demand data delivery specifically tailored for nearly-interactive vectorized analyses. It is capable of retrieving data from grid sites, on-the-fly data transformation, and delivering user-selected data in a variety of different formats. 
    New features will be presented that make the service ready for public use.
    An ongoing effort to integrate ServiceX with a popular statistical analysis framework in ATLAS will be described with an emphasis of a practical implementation of ServiceX into the physics analysis pipeline.

}
\maketitle
\section{Introduction}
\label{intro}
\begin{figure}[t]
\centering
\includegraphics[width=\textwidth]{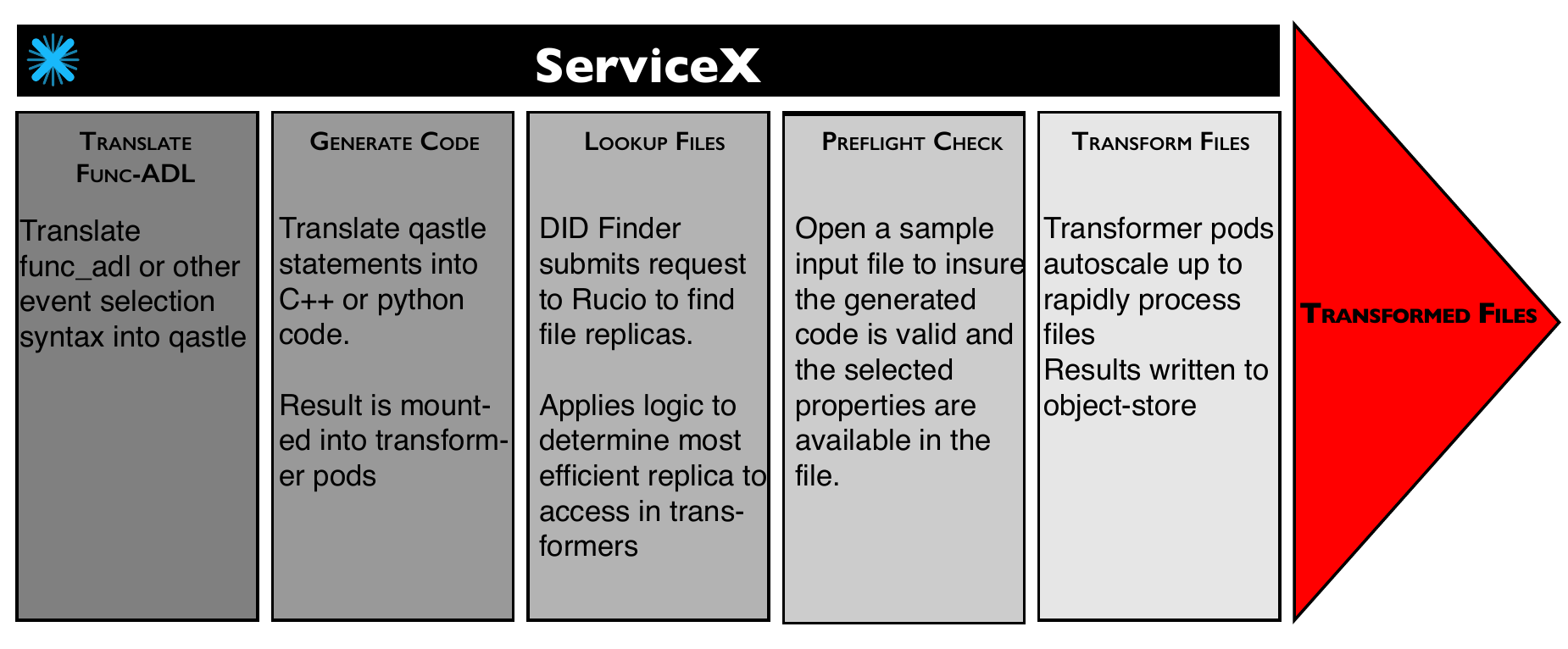}
\caption{End-to-end workflow of a ServiceX transformation run. The analysis description language, func-adl \cite{Ref-funcadl}, or ROOT TCut \cite{Ref-tcut} syntax are supported for a transformation query.}
\label{fig-servicex_value_chain}       
\end{figure}

ServiceX is a scalable HEP event data location, extraction, filtering, and transformation system that has been developed as part of the Institute for Research and Innovation in Software for High Energy Physics (IRIS-HEP). It runs on any Kubernetes cluster and can be offered as a public service or hosted on an institution's private cluster.

ServiceX accepts requests via a REST interface. The requests include a dataset identifier (DID) that resolves to a number of input data files along with a columnar event data selection statement expressed in an elemental expression language called Query AST Language Expressions (Qastle) \cite{Ref-qastle}. 
The service relies on Rucio \cite{Ref-rucio} to lookup file replicas for the requested DID. It will attempt to select the file replicas that are most efficiently accessible by the host Kubernetes cluster. 
The Qastle syntax is designed to be translated into code that describes operations on input data, with the transformer providing the data handling libraries. There is currently support for C++ code that is run in a transformer based on the ATLAS Event-Loop framework. Additionally, there is a python code generator that produces a script to drive the python Uproot \cite{Ref-uproot} library. It is suited for reading flat ntuples such as CMS NanoAOD \cite{Ref-nanoAOD} files and analysis group generated files.
The results from a ServiceX transformation are either flat ROOT files or Parquet columnar data files persisted to an object store which can be accessed via HTTP protocols or with the Amazon S3 API.
Figure~\ref{fig-servicex_value_chain} shows how each of these steps is assembled to produce the desired set of transformed files.

This paper introduces the latest developments in ServiceX that allow the system to be readily implemented into real-world applications.
Analysis pipelines that employ ServiceX for the currently available transformers are outlined.
In this paper, an analysis pipeline refers to a chain from the reconstructed object to the final result.
The primary goal of this paper follows to establish a practical implementation of ServiceX into the analysis pipeline that can be utilized in physics analysis.



\section{New features in ServiceX}
\label{sec-2}

ServiceX was first presented at CHEP 2019 \cite{Ref-chep19}. Since that conference, the system has evolved to make it useful as a production system to solve real world problems. The main enhancements were:
\subsection{Public access}
While ServiceX can be deployed in an experiment group's private Kubernetes cluster, users will still wish to connect to it from remote locations. As soon as the service is exposed to the internet, careful consideration must be given to securing it. In version 1.0, ServiceX uses Globus Auth \cite{Ref-globus} to authenticate users. Administrators are notified of user signups in a private Slack channel and can approve new accounts from there. Approved users are given an API token for making requests to the ServiceX deployment.

\subsection{Auto-scaling of Transformer Pods}
Initial versions of ServiceX required the user to specify the number of workers to launch to process the transformation job. This was simple to implement, but potentially wasteful in its use of resources, since CPUs could be sitting around idly. The service now makes use of Kubernetes auto-scaling capabilities to only launch new pods if files becoming available from Rucio exceed the existing set.

\subsection{Support for Reading ATLAS xAOD Files}
Reading of flat ntuples using Uproot libraries and Awkward arrays lends itself quite easily to the columnar style of analysis advanced by ServiceX. The ATLAS xAOD \cite{Ref-xaod} files are another matter all together as xAOD files use custom ROOT objects and require a C++ framework to fully utilize.  Research into the analysis description languages, func-adl, provided ServiceX with the ability to translate high level event selection queries into C++ code that is executed by an experiment approved framework inside the transformer pods. 

\subsection{Support for High-Level Expressions}
ServiceX uses an elemental LISP-inspired expression language called Qastle to represent event selection and transformation request. It is not intended for end-users to author selections in this representation. Instead, it is hoped that researchers of analysis description languages will create transpilers to generate Qastle queries from a higher level language. The ServiceX backend will immediately allow them experiment with their language using the full scale resources of the server. Initial work was done using the func-adl language from the Watts lab. As described in more detail in section \ref{sec:tcut}, it is equally important to work with popular event selection languages to encourage analysts to move their research over to this new environment with minimal disruption.






\section{ServiceX in analysis pipeline}
\label{sec-3}

\begin{figure}[b]
\centering
\includegraphics[width=\textwidth]{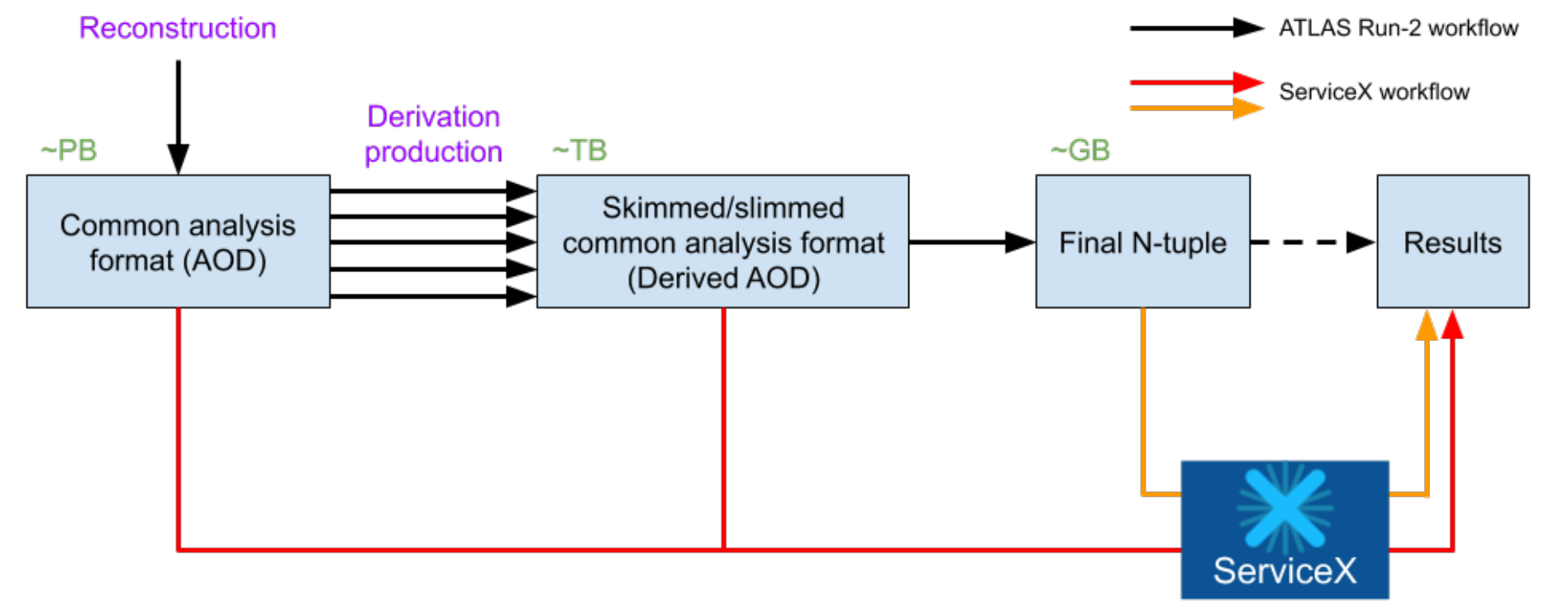}
\caption{A schema of ATLAS derivation production and subsequent analysis pipeline (black) and ServiceX pipelines (red/orange).}
\label{fig-servicex_analysis_pipeline}       
\end{figure}

Figure~\ref{fig-servicex_analysis_pipeline} shows a schema of the current ATLAS analysis pipeline starting from the AOD (Analysis Object Data) to the final analysis formats in black lines. Many individual and group-based derivations, which reduce the size of AOD by removing unnecessary information, exist and even smaller ROOT ntuples for the final analysis. 
ServiceX, on the other hand, can access directly to the upstream of the analysis pipeline as it supports different types of transformers for different input file formats.

The transformers available today are developed for those file formats that are primary in the analysis pipeline at present: xAOD transformer for the ATLAS xAOD or Derived AOD format (red lines), and Uproot transformer for flat ROOT ntuples (orange line). 
The latter is also compatible with the CMS NanoAOD format, which also features a flat ROOT TTree structure. A transformer for the CMS MiniAOD format, which also requires dedicated libraries similar to the ATLAS xAOD/DAOD format, is currently being developed.

The strong point of ServiceX is the flexibility of the transformer. The architecture implemented in ServiceX relies on a number of containerized microservices which include a transformer. Thus ServiceX is adaptable to future data formats with new transformers, and it is also feasible to have dedicated transformers for specific purposes.

\section{ServiceX for statistical analysis framework} 
\label{sec-4}

Given that ServiceX is relatively new to the community and the public release of the service became available recently, there are not many practical implementations of ServiceX into the analysis pipeline up to the present time.
It is also because of the fact that most analyzers are accustomed to the traditional analysis pipeline based on ROOT and grid jobs. Therefore, it is helpful to lower the barrier to allow them to experience the new analysis ecosystem in Python.
We have been developed a tool that can be implemented into a practical physics analysis with a minimum effort from analyzers.

\begin{figure}[b]
\centering
\includegraphics[width=\textwidth]{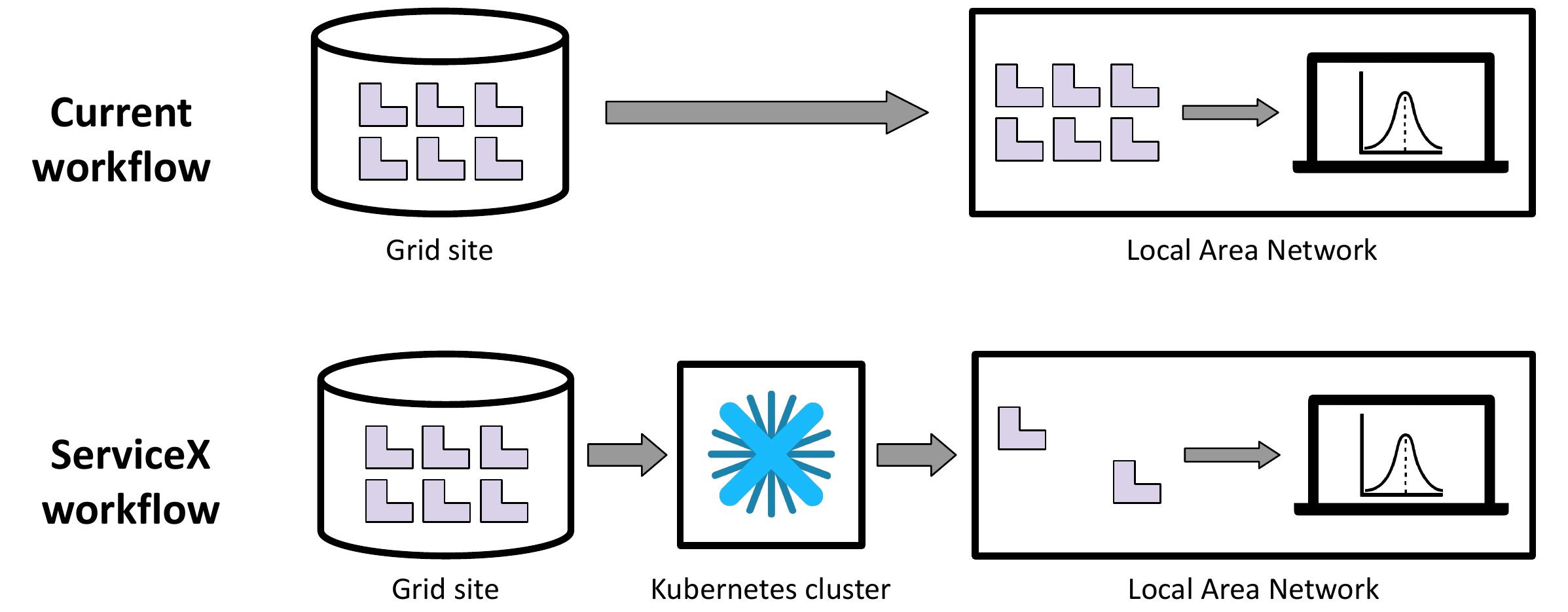}
\caption{Current workflow of TRExFitter (top) and alternative ServiceX workflow (bottom) to generate histograms from ROOT ntuples that are produced on the grid.}
\label{fig-trex_current_vs_servicex_workflow}       
\end{figure}

\subsection{TRExFitter}

TRExFitter \cite{Ref-trexfitter} is a framework used by many ATLAS physics analyses for statistical inference via profile likelihood fits.
It is designed to handle everything that analysers need for the statistical part of their analysis. It produces RooFit workspaces, perform fits on them, and interfaces with RooStats macros for limit and significance. It also generates publication-level pre-fit and post-fit plots and tables. Many more additional features are also supported to understand the fit behavior.

TRExFitter takes ROOT ntuples or histograms as input format, and a configuration file to steer the framework. A configuration file includes high-level physics choices: specification of signal/validation/control regions (the channels in HistFactory schema), observables to be used for the statistical analysis, Monte Carlo samples for signal and background and data samples, sources of systematic uncertainties, details of fit model, parameter of interest, and other general settings.

TRExFitter provides the feature to generate histograms for statistical analysis from ROOT ntuples based on the provided configuration file.
Hence, it is a typical workflow to download whole ROOT ntuples from the grid to a local cluster or directly accessible machines to generate histograms as shown in the top of Figure~\ref{fig-trex_current_vs_servicex_workflow}.
On the other hand, ServiceX can deliver only necessary branches or columns with event filtering as shown in the bottom of Figure~\ref{fig-trex_current_vs_servicex_workflow}.
The advantages of the ServiceX workflow are as follows:
\begin{itemize}
    \item Local storage: It takes up less storage space as ServiceX delivers minimal information to generate histograms.
    \item Download time: Data transferred over WAN is smaller for the ServiceX workflow. The network speed between a grid site and Kubernetes cluster is sufficiently fast as they are usually co-located.
\end{itemize}



\subsection{TCut translator for ServiceX}
\label{sec:tcut}

ServiceX utilizes func-adl, a python-based declarative analysis description language, to filter events and request branches from the input data file.
It is an intuitive language to extract data directly from ROOT ntuples, but TRExFitter relies on the ROOT \texttt{TTree::Draw} method, which uses TCut syntax for TTree selections.
Since TCut syntax is not directly readable by the func-adl, a python package for TCut to func-adl translation is developed. 
The package further converts func-adl to Qastle language, which is the language that ServiceX transformers understand.

The package supports arithmetic operators (\texttt{+, -, $\ast$, /}), logical operators (\texttt{!, $\&\&$, ||}), and relational and comparison operators (\texttt{==, !=, >, <, >=, <=}). Listing \ref{lst:tcut} shows an example of translating a ROOT-based query into a ServiceX query. The first argument is the name of the TTree object in an input file. The second is the list of selected branches for delivery. The last argument is the TCut object for TTree selection. Only events that pass the selection will be delivered for the requested branches. Thus, the second argument effectively removes branches from the input ROOT tree, and the third drops events. The package, {\small \texttt{tcut-to-qastle}} \cite{Ref-tcut-qastle}, is published at PyPI for a convenient access.

\begin{lstlisting}[language=Python, caption=An example translation of ROOT-based query into ServiceX query, label={lst:tcut} ]
import tcut_to_qastle as tq

# Get ServiceX query
query = tq.translate("nominal", "A,B,D", "(A && !B) || (C > 0.1)")
\end{lstlisting}

\subsection{$\texttt{servicex-for-trexfitter}$}
\label{subsec:servicex-for-trexfitter}

The {\small $\texttt{servicex-for-trexfitter}$} is a python package, which has been developed to provide seamless integration of ServiceX into the TRExFitter framework. It makes use of the building blocks that are described above: Uproot ServiceX to read input ROOT ntuples from the grid and perform transformations; ServiceX frontend library to access ServiceX backend and manage ServiceX delivery requests; the {\small \texttt{tcut-to-qastle}} package to translate ROOT-based query into the ServiceX query.

Each of the following steps runs within the {\small $\texttt{servicex-for-trexfitter}$} package: prepares ServiceX requests by analyzing a TRExFitter configuration file to deliver a minimum amount of data from the grid, makes ServiceX requests simultaneously, downloads output of finished transformations asynchronously from the object store of Kubernetes cluster, and converts downloaded output files into a ROOT file for each sample.

It takes a TRExFitter configuration file, which has an identical structure with the traditional ROOT ntuple input.
The only addition is a new field for grid dataset ID for each sample since ServiceX reads input ROOT ntuples from the grid.
The caching feature of the ServiceX frontend library allows only modified or added part of the TRExFitter configuration creates new ServiceX requests.

\subsection{Example}

The following prerequisites are needed to run the {\small \texttt{servicex-for-trexfitter}} package. 
It is written for Python version equal to or higher than 3.6. 
Access to an Uproot ServiceX endpoint is also required. 
Any running Uproot ServiceX instance should work, or access to the centrally-managed ServiceX instance can be granted as described in the ServiceX documentation \cite{Ref-servicex_doc}.
To convert the outputs from ServiceX into ROOT TTree, PyROOT \cite{Ref-pyroot} has to be installed.
Lastly, the input ROOT ntuples need to be organized in a way that each sample in the TRExFitter configuration file corresponds to a single Rucio dataset ID.

Listing \ref{lst:servicex_for_trexfitter} shows an example of how to use the {\small $\texttt{servicex-for-trexfitter}$}. An instance can be created with an argument of TRExFitter configuration file. 
Data transformation and delivery status can be interactively monitored just after the method {\small \texttt{get$\_$ntuples()}} is called, and the path to the slimmed/skimmed output ROOT ntuples will be printed once the delivery is completed.

\begin{lstlisting}[language=Python, caption=An example of running {\small $\texttt{servicex-for-trexfitter}$ to get slimmed/skimmed ROOT ntuples using ServiceX}, label={lst:servicex_for_trexfitter} ] 
from servicex_for_trexfitter import ServiceXTRExFitter
sx_trex = ServiceXTRExFitter("example.config")
sx_trex.get_ntuples()
\end{lstlisting}


\subsection{Benchmark results}

The performance of {\small $\texttt{servicex-for-trexfitter}$} is measured and compared with the current workflow using the same TRExFitter configuration file.
A practical TRExFitter configuration file which contains 17 Samples and 34 Systematics is used for the benchmark.
The total size of ROOT ntuples is 650 Gigabytes stored at the MidWest Tier-2 Center.
The benchmark for ServiceX workflow utilizes the Uproot ServiceX deployed at the University of Chicago SSL-River Kubernetes cluster.

\begin{table}[b]
\begin{center}
\begin{tabular}{ccc}
\hline
& Current workflow & {\small $\texttt{servicex-for-trexfitter}$} \\
\hline
Local disk space & 650 GB & 0.4 GB \\
Download ROOT ntuples & $>$hours & 7 mins \\
Process ROOT ntuples & 138 mins & 2.5 mins \\
\hline
\end{tabular}
\caption{Benchmark results from the current TRExFitter workflow and the workflow using {\small $\texttt{servicex-for-trexfitter}$}.}
\label{table:benchmark}
\end{center}
\end{table}

The results are shown in Table~\ref{table:benchmark}.
The current workflow takes up the same amount of local disk space with the total size of ROOT ntuples on the grid, whereas the workflow using {\small $\texttt{servicex-for-trexfitter}$} takes up a lot smaller disk space as it delivers only information that is needed to generate histograms defined in the TRExFitter configuration file.
The wall time to download ROOT ntuples for the subsequent step, generating histograms from downloaded ROOT ntuples, shows a much shorter time for the workflow using {\small $\texttt{servicex-for-trexfitter}$}.
This is due to the fact that ServiceX scales the workers to parallelize transformations and delivers only a subset of ROOT ntuples over WAN.
In addition, the time spent on processing downloaded ROOT ntuples is significantly shorter for the workflow using {\small $\texttt{servicex-for-trexfitter}$} as it needs to process only 0.4 GB than 650 GB.





\section{Conclusion and outlook}

The recent developments in ServiceX put the service in a state of a production system that allows practical implementations into a traditional analysis workflow. 
The authentication system which enables public access over the internet is particularly beneficial as it opens the door to more users.
The auto-scaling of transformer pods makes the system more robust, and the new C++ transformer for the ATLAS xAOD improves the performance significantly.
The support of TCut expressions to the func-adl language lowers the threshold to try the service.

The real-world application which employs the new features of ServiceX has been developed to provide a novel data delivery method to the popular statistical analysis framework in ATLAS.
Accessing input ROOT ntuples directly from the grid saves significant amount of local disk space.
Scaling up the transformer pods in a Kubernetes cluster can remarkably reduce a turnaround time by extracting only necessary information from input ROOT ntuples.
The primary goal of the {\small $\texttt{servicex-for-trexfitter}$} package is also fulfilled by requiring a minimal addition to the traditional approach.

This paper describes the implementation of ServiceX to interface with the existing ROOT-based statistical analysis framework. 
The HEP tools in Python are rapidly evolving, and ServiceX can nicely align with those tools to achieve a complete analysis workflow within the Python ecosystem.

%
%

\end{document}